\documentclass[twocolumn]{aastex6}
\usepackage{amsmath}

\fullcollaborationName{}

\def\simgt{\lower.5ex\hbox{\gtsima}} 
\def\simlt{\lower.5ex\hbox{\ltsima}} 
\def\gtsima{$\; \buildrel > \over \sim \;$} 
\def\ltsima{$\; \buildrel < \over \sim \;$}
\def\Msun{M_\odot}

\newcommand\lsim{\mathrel{\rlap{\lower4pt\hbox{\hskip1pt$\sim$}}
        \raise1pt\hbox{$<$}}}
\newcommand\gsim{\mathrel{\rlap{\lower4pt\hbox{\hskip1pt$\sim$}}
        \raise1pt\hbox{$>$}}}
\def\myputfigure#1#2#3#4#5%
{\vskip#5pt\makebox[0pt]{\hskip#2in
\includegraphics[width=#3\textwidth]{#1}}\vskip#4pt\hfill}

\begin{document}

\title{ Feedback Limits to Maximum Seed Masses of Black Holes }

%% Use \author, \affil, plus the \and command to format author and affiliation 
%% information.  If done correctly the peer review system will be able to
%% automatically put the author and affiliation information from the manuscript
%% and save the corresponding author the trouble of entering it by hand.
%%
%% The \affil should be used to document primary affiliations and the
%% \altaffil should be used for secondary affiliations, titles, or email.

\author{Fabio Pacucci\altaffilmark{1,2}, Priyamvada Natarajan\altaffilmark{1,3}, Andrea Ferrara\altaffilmark{2}}
\and

\altaffiltext{1}{Department of Physics, Yale University, New Haven, CT 06511, USA.}
\altaffiltext{2}{Scuola Normale Superiore, Piazza dei Cavalieri 7, I-56126 Pisa, Italy.}
\altaffiltext{3}{Department of Astronomy, Yale University, New Haven, CT 06511, USA.}

\begin{abstract}
The most massive black holes observed in the Universe weigh up to $\sim 10^{10} \, \mathrm{\Msun}$, nearly independent of redshift. Reaching these final masses likely required copious accretion and several major mergers. Employing a dynamical approach, that rests on the role played by a new, relevant physical scale - the transition radius - we provide a theoretical calculation of the maximum mass achievable by a black hole seed that forms in an isolated halo, one that scarcely merged. Incorporating effects at the transition radius and their impact on the evolution of accretion in isolated haloes we are able to obtain new limits for permitted growth. We find that large black hole seeds ($M_{\bullet} \gsim 10^4 \, \mathrm{\Msun}$) hosted in small isolated halos ($M_h \lesssim 10^9 \, \mathrm{\Msun}$) accreting with relatively small radiative efficiencies ($\epsilon \lesssim 0.1$) grow optimally in these circumstances. Moreover, we show that the standard $M_{\bullet}-\sigma$ relation observed at $z \sim 0$ cannot be established in isolated halos at high-$z$, but requires the occurrence of mergers. Since the average limiting mass of black holes formed at $z \gsim 10$ is in the range $10^{4-6} \, \mathrm{\Msun}$, we expect to observe them in local galaxies as intermediate-mass black holes, when hosted in the rare haloes that experienced only minor or no merging events. Such ancient black holes, formed in isolation with subsequent scant growth, could survive, almost unchanged, until present.
\end{abstract}


%%%%%%%%%%%%%%%%%%%%%%%%%%%%%%%%%%%%%%%%%%%%%%%%%%%%%%%%%%%%%%%%%%%%%%
%% SECTION 1: INTRODUCTION
%%%%%%%%%%%%%%%%%%%%%%%%%%%%%%%%%%%%%%%%%%%%%%%%%%%%%%%%%%%%%%%%%%%%%%
\section{Introduction}
\label{sec:introduction}
In recent years, extremely massive black holes, with masses of order $10^{10} \, \mathrm{\Msun}$, have been discovered both locally and in the early Universe. For instance, the local galaxy NGC 4889, harbors a Super-Massive Black Hole (SMBH) with mass $\sim 2 \times 10^{10} \, \mathrm{\Msun}$  \citep{McConnell_2011}.  Even higher masses, up to $\sim 4 \times 10^{10} \, \mathrm{\Msun}$, are estimated for blazars \citep{Ghisellini_2010}. Additionally, the highest redshift quasars detected at $z \sim 7$ also appear to be powered by SMBHs (e.g. \citealt{Fan_2001,Mortlock_2011, Marziani_2012, Wu_2015}). It is notable that these SMBHs in the nearby and far Universe have comparable masses, suggestive of an upper limit for black hole masses (\citealt{Natarajan_2009}, \citealt{Netzer_2003}).

The detection of SMBHs powering quasars at $z \sim 7$, the most massive highest redshift black holes, when the Universe was only $\sim 800 \, \mathrm{Myr}$ old, suggests that starting out with massive initial seeds could easily account for their growth histories. In fact, if the growth started from standard stellar-mass seeds ($\sim 10 \, \mathrm{\Msun}$) and was capped at the Eddington rate, the timing would be tight to produce the observed abundance of objects as massive as $\sim 10^{9-10} \, \mathrm{\Msun}$ already at $z \sim 7$ \citep{Lodato_Natarajan_2006, Volonteri_2010, Alexander_2014, Madau_2014, PVF_2015}. The observation that the most massive black holes ever discovered, both at $z \sim 0$ and $z \sim 7$, lie at masses $\sim 10^{10} \, \mathrm{\Msun}$ suggests that there is likely a physical mechanism that caps the growth around this value. Indeed, if $\gsim 10^{9} \, \mathrm{\Msun}$ black holes were in place at $\sim 7$, there would have been sufficient time to reach masses $M_{\bullet} \gg 10^{10}$ by $z \sim 0$. Currently, there is no observational evidence of \textit{accreting} objects with such extreme masses. This does not exclude the existence of non-accreting collapsed objects \citep{Natarajan_2009,King_2016}.

\subsection{Limits to black hole masses}

Several papers have investigated the physical mechanisms that might lead to an upper limit on the black hole masses. In the first discussion of this issue, \cite{Natarajan_2009} originally suggested that a limit $\sim 10^{10} \, \mathrm{\Msun}$ might be provided by self-regulation effects, arising from the co-evolution of the stellar component and the black hole harbored in the galactic nucleus. Several recent papers have revisited the argument: \cite{King_2016} shows that there is a physical limit to the mass of a black hole ($\sim 5 \times 10^{10} \, \mathrm{\Msun}$ for typical parameters) above which it cannot grow through luminous accretion of gas. This strict upper limit is obtained as for black hole masses above this limit an accretion disk cannot form, due to the fact that the radius of the innermost stable circular orbit would in such cases exceed the self-gravity radius (i.e where star formation is efficient), preventing the emission of radiation during the infall of gas. Meanwhile, \cite{Inayoshi_2016} find a similar limiting mass $\sim (1-6) \times 10^{10} \, \mathrm{\Msun}$, on the basis of the fate of the accreting gas. Growing more massive black holes would require an extremely large gas supply rate ($\gtrsim 10^3 \,\mathrm{\Msun \, yr^{-1}}$, assuming Eddington-limited accretion) which would be almost entirely converted into stars before reaching the black hole. All these arguments provide explanations for capping the masses on central black holes in galaxies. 

In this Letter, we investigate the limits that accretion physics on its own likely imposes on masses of black holes. This study utilized the concept of the transition radius (see Sec. \ref{sec:scales}), a spatial scale that helps connect and demarcate the large-scale dynamics of the accretion flow and the physics on smaller scales. Our work differs from previous studies as we investigate the maximum mass that a black hole seed can extract from the gas content of its parent halo, without interacting with other halos, i.e. for a high-$z$ black hole seed and its host halo that are isolated in the absence of mergers. The merger rate of high-$z$ galaxies is large and redshift dependent ($\sim (1+z)^{2.2-2.5}$, see e.g \citealt{Rodriguez-Gomez_2015}) and it can reach values of several events per Gyr at $z \gsim 6$. Nonetheless, the motivation of our work is two-fold. First, the accretion time scale of high-$z$ black hole seeds is generally shorter than the merger rate, topping at a few $\sim 100 \, \mathrm{Myr}$: therefore our work provides a typical value for the black hole mass prior to the occurrence of mergers. Second, in the unlikely but still plausible event that a halo has practically no major mergers, we should be able to find at $z \sim 0$ black holes with masses similar to the maximum value computed here. In this calculation, we assume that the black hole seed forms at (or slowly migrates to) the center of its host halo, irrespective of the specific formation channel for the black hole seed.

%%%%%%%%%%%%%%%%%%%%%%%%%%%%%%%%%%%%%%%%%%%%%%%%%%%%%%%%%%%%%%%%%%%%%%
%% SECTION 2: OUR MODEL
%%%%%%%%%%%%%%%%%%%%%%%%%%%%%%%%%%%%%%%%%%%%%%%%%%%%%%%%%%%%%%%%%%%%%%
\section{Limiting mass from accretion physics}
\label{sec:scales}

\begin{figure}
\vspace{-1\baselineskip}
\hspace{-0.5cm}
\begin{center}
\includegraphics[angle=0,width=0.5\textwidth]{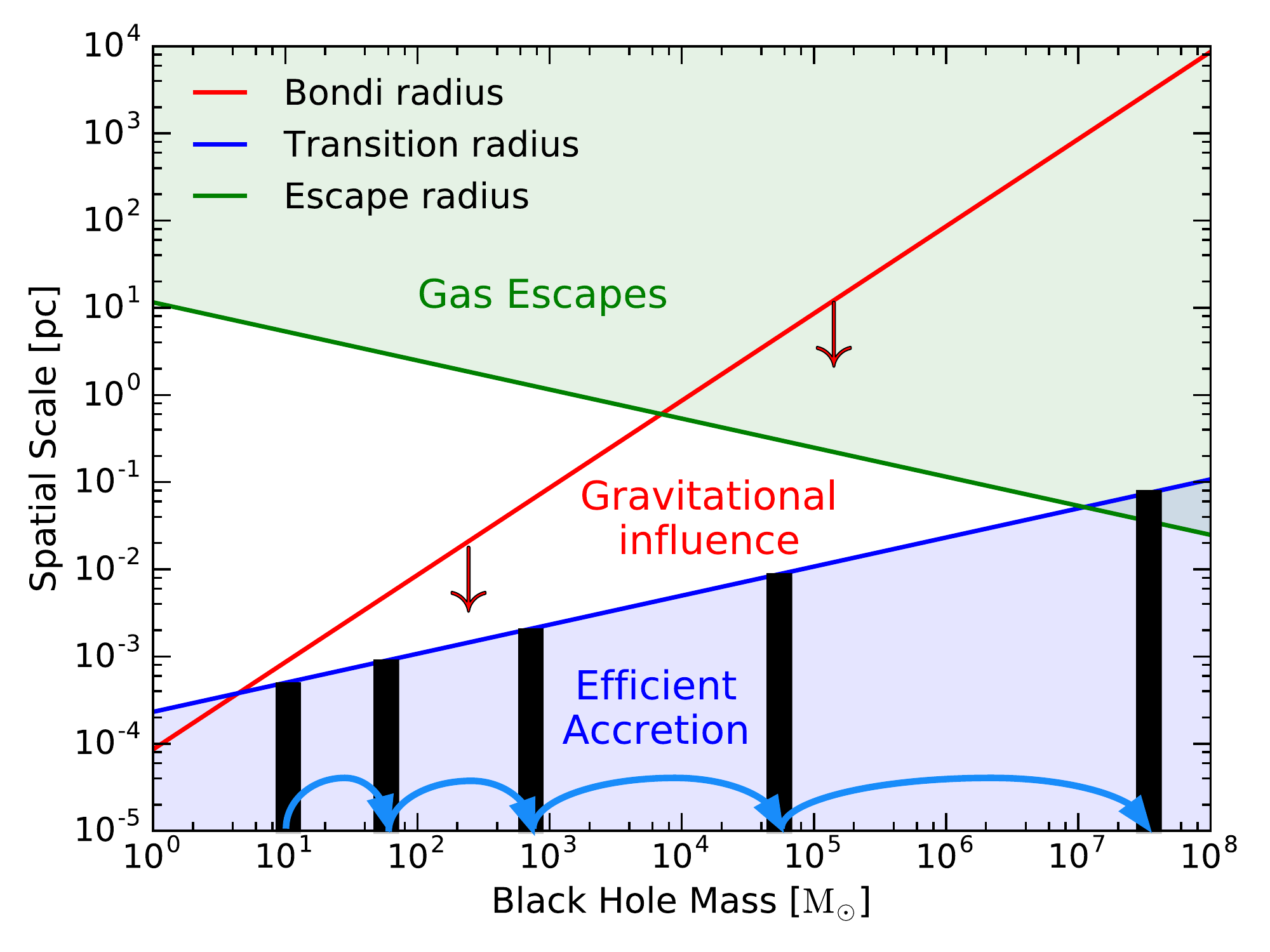}
\caption{Summary of the spatial scales relevant to this work, as a function of $M_{\bullet}$ ($\rho_0 \sim 10^{-18} \, \mathrm{g \, cm^{-3}}$, $a \sim 1 \, \mathrm{pc}$). The Bondi radius is shown in red. On scales smaller than the transition radius (blue line), accretion is rapid. On scales larger than the escape radius (green line), the gas is lost through outflows. The rapidity of the black hole growth depends on the initial mass of the seed: more massive objects grow more rapidly (see Sec. \ref{subsec_accretion_easier} for further details).}
\label{fig:scales}
\end{center}
\end{figure}

Here we focus on the limits imposed on the growth of central black holes from accretion physics in isolated host halos. Assuming that the black hole seed has initial mass $M_{\bullet 0}$, the estimate of the maximum mass that can be accreted follows from the comparison of different spatial scales, described below. To facilitate comparison, we scale them in units of the Schwarzschild radius:
\begin{equation}
R_S = \frac{2 G M_{\bullet}}{c^2} \approx 3 \times 10^5 \left( \frac{M_{\bullet}}{10^5 \, \mathrm{\Msun}} \right) \, \mathrm{km} \, ,
\end{equation}
where $G$ is the gravitational constant and $c$ the speed of light. Additionally, we assume an isothermal density profile for the gas content of the host halo:
\begin{equation}
\rho(r) = \frac{\rho_0}{1+(r/a)^2} \, ,
\label{eq:rho_def}
\end{equation}
In the example of Fig. \ref{fig:scales}, $\rho_0 \sim 10^{-18} \, \mathrm{g \, cm^{-3}}$ and $a \sim 1 \, \mathrm{pc}$. Moreover, the gas distribution is extended up to $\sim 100 \, \mathrm{pc}$.
A visual summary of the various relevant spatial scales, as a function of $M_{\bullet 0}$, is shown in Fig. \ref{fig:scales}.

\subsection{The Bondi radius}

Traditionally the Bondi radius has served as the typical radius from within which all gas enclosed can, in principle, be accreted.
Following \cite{Bondi_1952}, we define it as:
\begin{equation}
R_B = \frac{2 G M_{\bullet}}{c_s^2} \, ,
\label{r_g_definition}
\end{equation}
where $c_s$ is the sound speed of the unperturbed gas surrounding the black hole. In terms of $R_S$, this can be re-written as:
\begin{equation}
R_B \approx 9 \times 10^8 \, R_S \, \left( \frac{c_s}{10 \, \mathrm{km \, s^{-1}}} \right)^{-2} \, .
\end{equation}
Inside $R_B$, the gas content of the halo is perturbed by the presence of the black hole. Consequently, at a given time $t$, the Bondi radius of a black hole contains the maximum amount of mass that can be accreted. The sphere of gravitational influence expands with time, as $\propto M_{\bullet}$.

\subsection{The transition radius}

Another spatial scale, the transition radius (defined by \citealt{PVF_2015}), demarcates the outflow dominated region from the inflow dominated one. It follows from the comparison of two time scales, the typical feedback time-scale ($t_{fb}$) and the accretion time ($t_{acc}$). The former, computed at some radius $r$, is the time needed by the radiation pressure to significantly (i.e. by a factor $e$) alter $\dot{M}_{\bullet}$. The latter estimates the time needed to accrete the gas mass inside a sphere of radius $r$. The transition radius is defined as the radius where $t_{fb}(R_T) = t_{acc}(R_T)$:
\begin{equation}
R_{T} = \left[ \psi \frac{3Gf_{Edd}}{(\epsilon \rho_0 \kappa_e c)^2} M_{\bullet} \right]^{1/3} \, ,
\label{r_t_definition}
\end{equation}
where $f_{Edd} \equiv \dot{M_{\bullet}}/\dot{M}_{Edd}$ is the ratio of the accretion rate to the Eddington rate; $\epsilon$ is the matter-radiation conversion factor; $\kappa_e$ is the electron scattering opacity and $\psi$ is a numerical factor that depends on the details of the adopted accretion model. In the standard radiatively efficient scenario $f_{Edd} = 1$,  $\epsilon = 0.1$ and $\psi =1$, while in highly-obscured environments $f_{Edd} >1$, $\epsilon \lsim 0.04$ and $\psi \gsim 25$. Written out in terms of $R_S$, folding in typical values for the parameters above, we have:
\begin{align}
R_T \approx & 10^6 \, R_S (\psi f_{Edd})^{1/3} \left( \frac{\epsilon}{0.1} \right)^{-2/3} \times \\
\nonumber & \times \left( \frac{\rho_0}{10^{-18} \, \mathrm{g \, cm^{-3}}} \right)^{-2/3} \left( \frac{M_{\bullet}}{10^5 \, \mathrm{\Msun}} \right)^{-2/3} \, .
\end{align}

For $r \gtrsim R_T$ the radiative feedback dominates over gravity, while for $r \lesssim R_T$ accretion is rapid. The transition radius is then the spatial scale above which the accretion flow is dominated by the radiation pressure that powers the appearance of outflows. The definition of $R_T$ allows us to define two accretion regimes \citep{PVF_2015, Park_2016}: a feedback-limited growth, when radiative feedback is important, and a supply-limited one, when it is unimportant and most of the available gas is accreted.

\subsection{The escape radius}
\label{subsec:R_E} 

The one other crucial spatial scale is the escape radius beyond which gas remains unbound. If a gas particle is at $r \gtrsim R_T$ then it moves away from the black hole, but this does not necessarily imply that it leaves its gravitational field. If the accretion onto the central black hole is intermittent \citep{Pacucci_2015}, then the radiative acceleration on the particle is not continuous and it may never be able to reach the escape velocity from the halo.  Assuming intermittent accretion, the escape radius $R_E$ defines the distance from the black hole out of which a gas particle has a velocity larger than the escape velocity $v_e$ of the halo:
\begin{equation}
v_e = \sqrt{\frac{2 G M_h}{R_E}} \, ,
\end{equation}
where $M_h$ is the total mass of the halo (baryons and dark matter). The velocity of the outflow at the transition radius may be computed as:
\begin{equation}
v_o = \frac{L_{edd}}{\dot{M} c} = \frac{4 \pi G M_{\bullet}}{\kappa_e \dot{M}} \, ,
\end{equation}
where the mass flux $\dot{M}$ is: $\dot{M}(R_T) = 4 \pi R_T^2 \rho(R_T) v_o$. We assume that the outflow is momentum-driven instead of energy-driven because the process is not adiabatic 
(\citealt{Pacucci_2015}). In the outflow region ($r \gtrsim R_T$), the gas is lost if $v_o \gsim v_e$:
\begin{equation}
R_E = 2 \left( \frac{M_h}{M_{\bullet}} \right) \kappa_e R_T^2 \rho(R_T) \, .
\label{R_E_definition}
\end{equation}
If $R_E \lesssim R_T$ the definition of the escape radius is meaningless, since inside $R_T$ all the gas is accreted. Requiring $R_E/R_T > 1$ leads to:
\begin{equation}
\left( \frac{M_h}{M_{\bullet}} \right) \kappa_e R_T \rho(R_T) > 0.5 \, .
\end{equation}
The optical depth of the gas, computed at $R_T$ and modulated by the ratio $M_h/M_{\bullet}$, needs to be of order unity to have mass loss from the halo.
Expressing $R_E $ in terms of $R_S$:
\begin{align}
R_E \approx & 2 \times 10^7 \, R_S \left( \frac{M_h}{10^8 \, \mathrm{\Msun}} \right) \left( \frac{\rho(R_T)}{10^{-18} \, \mathrm{g \, cm^{-3}}} \right)  (\psi f_{Edd})^{2/3} \times  \\ 
\nonumber & \times  \left( \frac{\epsilon}{0.1} \right)^{-4/3} \left( \frac{\rho_0}{10^{-18} \, \mathrm{g \, cm^{-3}}} \right)^{-4/3} \left( \frac{M_{\bullet}}{10^5 \, \mathrm{\Msun}} \right)^{-4/3} \, .
\end{align}
The spatial scales introduced so far in Sec. \ref{sec:scales} can now be used to investigate the main features of the accretion flow.

\section{RESULTS}
\label{sec:results}

\subsection{Outflows are inevitable for realistic density profiles}

It is only under the condition $R_T>R_B$ that there are no outflows since all the gas affected by the gravitational field of the black hole is promptly accreted. To obtain the maximum growth possible, when all the available gas is accreted, the ideal situation would therefore be to have $R_T \gsim R_B$ always. This condition, however, is not easily satisfied. Equating the definitions of $R_B$ (Eq. \ref{r_g_definition}) and $R_T$ (Eq. \ref{r_t_definition}), we derive that $R_T= R_B$ for  $M_{\bullet} \approx 4 \, \mathrm{\Msun}$ for our adopted standard values for $\rho_0$ and $a$. Since this mass is tiny (just above the Oppenheimer-Volkoff limit of $\sim 3 \, \mathrm{\Msun}$ to form a black hole) and since $R_T \propto M_{\bullet}^{1/3}$ while $R_B \propto M_{\bullet}$, we obtain that $R_B > R_T$ always. Therefore, for realistic density profiles, the occurrence of outflows is inevitable. For very low density conditions ($\lesssim 10^{-20} \, \mathrm{g \, cm^{-3}}$, a central density $\sim100$ times lower than our standard value), the mass that gives $R_T = R_B$ is larger ($\gsim 100 \, \mathrm{\Msun}$) and the central black hole may be able to promptly accrete all the gas inside $R_B$ (see Fig. \ref{fig:m_h}). However, given the very low value of the gas mass available, the black hole would not be able to grow substantially before merging with other halos.

\subsection{Accretion is easier for larger black hole masses}
\label{subsec_accretion_easier}

The transition radius scales as $R_T \propto M_{\bullet}^{1/3}$ while the escape radius scales as $R_E \propto M_{\bullet}^{-1/3}$. Therefore, for increasing $M_{\bullet}$ the fraction of gas mass available for prompt accretion increases, as does the amount of outflowing gas. Nonetheless, this trend reaches a turning point at $M_{\mathrm{crit}}$ where the limit $R_E=R_T$ is reached (see Fig. \ref{fig:scales} where $M_{\mathrm{crit}} \sim 10^{7} \, \mathrm{\Msun}$). For $M_{\bullet} \gtrsim M_{\rm crit}$ the definition of $R_E$ ceases to be valid, because all the gas inside the transition radius is promptly available for accretion.
This result can be summarized as follows: \textit{accretion is easier for black hole seeds with larger masses} (see \citealt{PVF_2015, Inayoshi_2015, Park_2016}). Consequently, the subsequent rapidity of growth of a black hole seed depends on its formation as that determines its initial mass.

Fig. \ref{fig:scales} visually explains the concept. Assuming that a black hole seed starts to grow from $\sim 10 \, \mathrm{\Msun}$, an idealized example of the growth path is shown with cyan arrows. The initial seed is able to rapidly accrete mass within and up to its transition radius, which leads to a small increase in mass (and consequently in $R_T$), shown by the first arrow. At the next step, the black hole is able to accrete a larger amount of mass rapidly. This mechanism leads to increasingly larger steps in the mass growth, implying that the larger the initial mass of the seed, the faster it can grow to higher masses. While the accretion rate is not necessarily proportional to the transition radius, the amount of gas available for prompt accretion (i.e. without the radiation pressure interfering) scales with $R_T$.

\subsection{Maximum mass in isolated halos}
\label{sec:maximum_mass}

With the relevant physical scales defined, we can now ask, starting from a high-$z$ black hole seed of mass $M_{\bullet 0}$, formed in a halo of initial mass $M_{\rm h 0}$ and initial gas mass $M_{\rm g 0} \equiv f_b  M_{\rm h 0} $, what is the maximum mass that the central black hole can reach in the absence of interaction with other halos. Defining $\dot{M_{\bullet}}$ as the accretion rate onto the black hole and $\dot{M}_{out}$ the mass flux of gas located outside $R_E$, the following equations hold:
\begin{equation}
M_{\bullet}(t) = M_{\bullet 0} + \int_0^t \dot{M_{\bullet}}(t) \, {\rm dt} \, .
\end{equation}
\begin{equation}
M_{\rm g}(t) = M_{\rm g 0} - \int_0^t \dot{M}_{out}(t) \, {\rm dt} - \int_0^t \dot{M_{\bullet}}(t) \, {\rm dt} \, .
\end{equation}
The maximum black hole mass is reached at a time $t_m$ when the gas mass inside the host halo is completely depleted, $M_{\rm g}(t_m) = 0$, leading to the following equation for the maximum black hole mass $M_{\rm BH,max}$:
\begin{equation}
M_{\rm BH,max} = M_{\bullet 0} + M_{\rm g 0} - \int_{0}^{t_m} \dot{M}_{out}(t) \, {\rm dt} \, .
\label{eq:max_mass}
\end{equation}
The ejected mass flux can be computed as:
\begin{equation}
\dot{M}_{out}(t) = 4 \pi R_E(t)^2 \rho(R_E) v(R_E) \, ,
\end{equation}
and since it is time dependent: the solution of Eq. \ref{eq:max_mass} requires a numerical computation.

Nonetheless, it is possible to devise a simple procedure to obtain an approximate value for $M_{\rm BH,max}$. We showed that $R_E \propto M_{\bullet}^{-1/3}$ while $R_T \propto M_{\bullet}^{1/3}$ until $R_T = R_E$ at $M_{\bullet} = M_{\rm crit}$ and from this time on, the gas is unable to leave the halo. We assume that $M_{\rm crit}$ is also the maximum black hole mass achievable within an isolated halo. In fact, the time needed to accrete all the gas mass inside some given radius scales as $M_{\bullet}^{-1}$ (i.e. the accretion time scale, see \citealt{PVF_2015}); similarly, since $R_T \propto M_{\bullet}^{1/3}$, then the time scale at which the gas mass inside $R_T$ increases is also $\propto M_{\bullet}^{-1}$. In other words, the black hole accretes the gas inside its transition radius at the same pace that the transition radius grows. Similarly, given the scalings, outflows remove the gas outside the escape radius at the same pace at which the outflow zone expands. Hence, by the time the black hole mass reaches the value $M_{\rm crit}$, the gas mass inside the host halo has been completely ejected or accreted. Therefore, regardless of the fate of the gas, the halo is entirely depleted, thereby providing us the limiting value for the black hole mass.
The maximum black hole mass is then equal to $M_{\rm crit} (R_T = R_E)$ and can be computed as:
\begin{equation}
M_{\rm BH,max} = \left[ 3\psi G f_{Edd} \kappa_e \frac{(2 M_h \rho(R_T))^3}{(\epsilon \rho_0 c)^2} \right]^{1/2} \, .
\label{max_mass}
\end{equation}
Defining a mass scale:
\begin{equation}
M_{*} = \frac{M_{\bullet}}{[2R_T \kappa_e \rho(R_T)]^3} \, ,
\end{equation}
Eq. \ref{max_mass} can be written in a compact form as:
\begin{equation}
M_{\rm BH,max} = \left( \frac{M_h^3}{M_{*}} \right)^{1/2} \, .
\end{equation}

\begin{figure}
\vspace{-1\baselineskip}
\hspace{-0.5cm}
\begin{center}
\includegraphics[angle=0,width=0.5\textwidth]{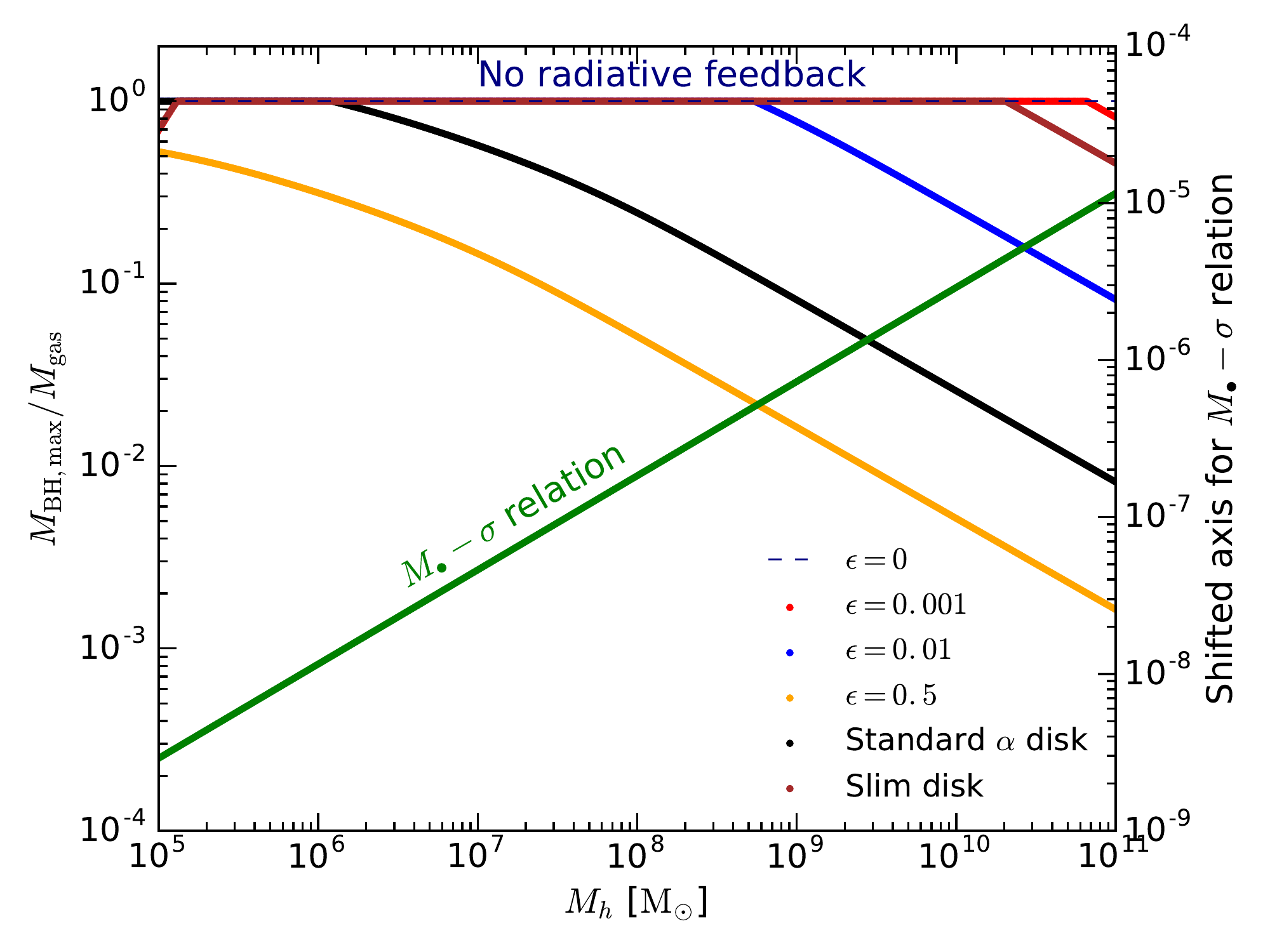}
\caption{Maximum black hole mass (rescaled by the gas mass of the halo) achievable as a function of the total mass of the host halo, for different adopted values for the radiative efficiency of the accretion flow. The case $\epsilon = 0$ corresponds to the absence of radiative efficiency, thus all the gas mass is accreted. The standard $\alpha$ disk and slim disk cases \citep{Abramowicz_1988,PVF_2015} are also shown. The green line (whose mass scale is on the right vertical axis) shows the $M_{\bullet}-\sigma$ relation, computed assuming an isothermal density profile.}
\label{fig:m_h}
\end{center}
\end{figure}

Fig. \ref{fig:m_h} shows the maximum black hole mass ($M_{\rm BH, max}$) as a function of the total mass (baryons + dark matter) of the host halo, for several adopted prescriptions for the radiative efficiency of the accretion flow. In the plot, $M_{\rm BH, max}$ is rescaled to the gas mass $M_{\rm gas}$ of the halo, so that a unit ratio means that all the available gas is accreted. The different halo masses explored in the figure are obtained by modifying $\rho_0$ and $a$ for the gas density distribution, and then rescaling it with the baryon fraction $f_b=0.1$.  We note that this calculation is redshift-independent, since it assumes that halos of any mass are available.
While, in general, a decrease in the radiative efficiency leads to an increase in the accretion rate, the actual scaling between $\epsilon$ and the mass growth is far from being trivial: in Fig. \ref{fig:m_h} we show examples of this relation for several relevant cases.

Interestingly, we note that from Fig. \ref{fig:m_h} it is apparent that the local $M_{\bullet}-\sigma$ relation cannot be created in isolated halos at high-$z$, but requires the occurrence of several merging events for establishment. Within isolated halos in the high-$z$ Universe, central black holes are hugely over-massive compared to their counterparts as seen in the local $M_{\bullet}-\sigma$ relation. While in the local Universe larger halos are generally associated with more massive black holes, in isolated high-$z$ halos the fraction of gas mass available for black hole growth decreases above a given threshold of the halo mass. In fact, a key insight from this approach to modeling is that very large values of $M_h$ increase the propensity for outflows to occur (see the definition of $R_E$, Eq. \ref{R_E_definition}) and therefore modulate the mass of the growing central black hole.

%%%%%%%%%%%%%%%%%%%%%%%%%%%%%%%%%%%%%%%%%%%%%%%%%%%%%%%%%%%%%%%%%%%%%%
%% SECTION 5: Discussion and Conclusions
%%%%%%%%%%%%%%%%%%%%%%%%%%%%%%%%%%%%%%%%%%%%%%%%%%%%%%%%%%%%%%%%%%%%%%
\section{Discussion and Conclusions}
\label{sec:disc_concl}

The most massive observable black holes in the Universe weigh up to $\sim 10^{10} \, \mathrm{\Msun}$, nearly independent of redshift. As per our current standard picture, we believe that these cosmic behemoths accreted gas and also merged with several halos to reach these final masses. In this Letter, we calculate the maximum mass achievable by a black hole seed that forms and grows in an isolated halo, using simple scaling arguments from relevant spatial scales and accretion physics. 

In the context of a fairly standard set of assumptions, we show that the maximum black hole mass reachable in an isolated halo depends on: (i) the total mass of the host halo, and (ii) the radiative efficiency 
of the accretion flow. Furthermore, we find that the rapidity of black hole growth depends on the initial mass of the black hole seed. Therefore, we conclude that large black hole seeds ($M_{\bullet} \gsim 10^4 \, \mathrm{\Msun}$) hosted in small isolated halos ($M_h \lesssim 10^9 \, \mathrm{\Msun}$) accreting with relatively small radiative efficiencies ($\epsilon \lesssim 0.1$) grow optimally. 

Additional effects (e.g. star formation, supernova explosions, non-negligible angular momentum of the gas) are not included in the present purely dynamical treatment. All these processes tend to reduce the effective amount of gas available in the overall reservoir for black hole growth. For this reason, our model provides an \textit{upper limit to the black hole mass} that can be grown in isolated halos, assuming that all these effects are negligible. In the environmental conditions available in the early Universe, though, we do predict these effects to be negligible for the following reasons: (i) the bulk angular momentum of high-$z$, isolated halos, which are the main objects of this model, is very small \citep{deSouza_2013}; (ii) relevant increases of the bulk angular momentum are reached after major mergers events, with external halos providing the required torque: and such events are discarded in our treatment; and (iii) major star formation activity also appears to be triggered by merger events which are beyond our purview in this treatment of evolution of black hole growth in isolated environments.

Finally, we show that the empirically derived $M_{\bullet}-\sigma$ relation observed at $z \sim 0$ cannot be established in isolated halos at high-$z$, as it requires the occurrence of several merging events. Accretion physics constraints show that while mergers are required to explain the SMBHs seen locally, the SMBHs at higher redshift can assemble with rapid gas accretion even in very massive and single, isolated halos. One important potential application of our methodology is that it offers a robust, physically motivated prescription to statistically populate high-$z$ halos in numerical simulations with reasonable initial black hole masses to follow subsequent black hole growth. Our picture also confirms that if a high-$z$ black hole seed formed in a halo that scarcely merged throughout cosmic history until $z=0$, then its mass should be observable now in the intermediate-mass black hole range ($10^{4-6} \, \mathrm{\Msun}$), as already pointed out by \cite{Volonteri_2008}, \cite{vanWassenhove_2010} and \cite{Habouzit_2016}. Indeed, assuming a standard $\alpha$ disk accretion flow (black line in Fig. \ref{fig:m_h}), halos with a reasonable probability to form at $z \gsim 10$ (i.e. $M_h < 10^{9} \, \mathrm{\Msun}$) will contain intermediate-mass black holes today. And for the rare haloes that have had scant mergers over cosmic time, this initial seeding and growth might be reflected in the black hole masses that they harbor today (e.g. \citealt{Davis_2011, Cseh_2014}).

\vspace{0.2cm}
F.P. acknowledges the SAO Chandra grant No. AR6-17017B and NASA-ADAP grant No. MA160009.
P.N. acknowledges support from a Theoretical and Computational Astrophysics Networks grant No. 1332858 from the NSF.

%%%%%%%%%%%%%%%%%%%%%%%%%%%%%%%%%%%%%%%%%%%%%%%%%%%%%%%%%%%%%%%%%%%%%%
%% BIBLIOGRAPHY
%%%%%%%%%%%%%%%%%%%%%%%%%%%%%%%%%%%%%%%%%%%%%%%%%%%%%%%%%%%%%%%%%%%%%%

\end{document}